\documentclass[%
 reprint,
superscriptaddress,
 amsmath,amssymb,
 aps,
]{revtex4-1}

\usepackage{booktabs}
\usepackage{multirow}

\usepackage{subcaption}

\usepackage{bm} 
\usepackage{graphicx}% Include figure files
\usepackage{dcolumn}% Align table columns on decimal point
\usepackage{bm}% bold math
\usepackage{hyperref}% add hypertext capabilities
\begin{document}

\preprint{APS/123-QED}

\title{Periodic Orbits and Gravitational Radiation from Extreme Mass-Ratio Inspirals as Probes of Black Hole Quantum Hair}% Force line breaks with \\
\author{Yiru Zhang}

\author{Meirong Tang}
%\email{tangmr@gzu.edu.cn(Corresponding author)}
%\affiliation{College of Physics, Guizhou University, Guiyang 550025, China}

\author{Zhaoyi Xu}%
\email{zyxu@gzu.edu.cn(Corresponding author)}
\affiliation{%
 College of Physics,Guizhou University,Guiyang,550025,China
}%

%\date{\today}% It is always \today, today,
             %  but any date may be explicitly specified

\begin{abstract}

The classical no-hair theorem states that stationary black holes in general relativity can be completely described by only a small set of global parameters. Within this framework, no additional geometric structures are expected to persist outside the event horizon. However, quantum vacuum polarization may introduce small modifications to the near-horizon geometry, effectively giving rise to what is known as quantum hair. Such corrections may provide a possible window into the microscopic structure and thermodynamic properties of black holes.
In this work, we examine how the quantum hair parameter $\gamma$ influences the periodic orbital dynamics of test bodies in extreme mass-ratio inspirals (EMRIs) and their associated gravitational-wave emission. We find that $\gamma$ significantly modifies the characteristic radii and angular momenta of two important circular orbits, namely the marginally bound orbit (MBO) and the innermost stable circular orbit (ISCO), leading to a shift in the allowed region of the energy--angular momentum ($E$--$L$) phase space. Based on the rational number $q$ classification, we further show that quantum corrections tend to enhance the zoom--whirl orbital behavior.Gravitational-wave calculations using the Numerical Kludge approach indicate that quantum hair alters the effective spacetime potential, produces small drifts in the fundamental orbital frequencies, and consequently leads to observable phase dephasing in long-duration signals. These results provide a dynamical signature for distinguishing quantum-corrected black holes from classical Schwarzschild ones and offer theoretical motivation for testing quantum gravity effects with future space-based gravitational-wave observatories.

\begin{description}
\item[Keywords]
quantum hair; quantum-corrected black holes; periodic orbits; extreme mass ratio inspirals; gravitational waves
%\item[Structure] 
%You may use the \texttt{description} environment to structure your abstract;
%use the optional argument of the \verb+\item+ command to give the category of each item.
\end{description}
\end{abstract}

%\keywords{keywords}%Use showkeys class option if keyword
                             % display desired
%Scalar field; Rotating black-bounce black hole;  Weak cosmic censorship conjecture; Spacetime singularity.
\maketitle

%\tableofcontents

\section{\label{sec:level1}Introduction}

General Relativity provides the fundamental theoretical framework for describing gravitational interactions in the classical regime. Over the past century, it has achieved remarkable success in explaining a wide range of astrophysical and cosmological phenomena, from precision measurements within the Solar System to the recent direct observations of gravitational waves. Within this theoretical framework, black holes formed through gravitational collapse are described by the Kerr–Newman class of solutions, where rotation and electric charge are encoded in the Kerr metric and its Newman generalization, respectively (e.g.,\cite{Wald:1984rg,Newman:1965my}). In addition, the black hole uniqueness theorems further indicate that, under the assumption of asymptotic flatness, stationary solutions can be completely characterized by only a few global conserved quantities. (e.g.,\cite{Israel:1967wq,Carter:1971zc,Robinson:1975bv}). These results lead to the well-known no-hair theorem. According to this theorem, a stationary black hole in asymptotically flat spacetime is completely specified by three macroscopic parameters: the mass $M$, angular momentum $L$, and electric charge $Q$. No additional geometric degrees of freedom remain outside the event horizon \cite{Chrusciel:2012jk}. Within this classical picture, any detailed microscopic information associated with the initial matter configuration—such as baryon or lepton number and possible inhomogeneities—is expected to be carried away through gravitational radiation during the collapse, leaving a smooth and nearly featureless event horizon. \cite{Price:1971fb}.

However, this classical geometric description faces fundamental challenges once quantum effects are included. In 1975, Hawking demonstrated that quantum field theory in curved spacetime predicts that black holes emit thermal radiation, now known as Hawking radiation, which eventually leads to black hole evaporation \cite{Hawking:1975vcx}. If this radiation were exactly thermal and carried no correlations with the internal degrees of freedom of the black hole, the complete evaporation process would transform an initially pure quantum state into a mixed state. Such a transition would violate the principle of unitary evolution in quantum mechanics and thereby gives rise to the black hole information paradox (e.g.,~\cite{Hawking:1976ra,Mathur:2009hf}). The AMPS firewall proposal further argues that maintaining unitarity near the event horizon may require significant modifications of the local energy structure, potentially conflicting with the equivalence principle of general relativity \cite{Almheiri:2012rt}. These considerations suggest that the classical picture of black holes may need to be revised by quantum gravitational effects in the near-horizon region or at Planckian scales.

In recent years, several candidate theories of quantum gravity have suggested that the conventional horizon picture may break down and that black holes could possess additional information-carrying structures often referred to as quantum hair. For example, within string theory, the fuzzball proposal developed by Mathur and collaborators proposes that the classical horizon is replaced by a finite-size configuration composed of strings and branes, thereby avoiding both singularities and information loss (e.g.,\cite{Mathur:2005zp,Skenderis:2008qn}). Another possibility is the soft hair scenario proposed by Hawking, Perry, and Strominger, in which soft photons or gravitons at null infinity introduce additional degrees of freedom on the horizon through supertranslation symmetries \cite{Hawking:2016msc}. Loop Quantum Gravity also suggests that spacetime discreteness may induce non-perturbative quantum corrections to the horizon through geometric fluctuations (e.g.,\cite{Ashtekar:2005cj,Rovelli:1996dv,Ashtekar:1997yu,Perez:2017cmj}). These developments collectively point toward the existence of quantum structures encoding black hole microstates that may manifest outside the classical horizon through gravitational modifications.

Despite these theoretical advances, direct observation of Planck-scale quantum effects remains extremely challenging. Effective Field Theory (EFT) provides a model-independent framework to bridge microscopic physics and macroscopic observables (e.g.,\cite{Burgess:2003jk,Donoghue:1994dn}). Within EFT, quantum gravitational effects can be described through higher-curvature corrections to the Einstein–Hilbert action. Pioneering work by Bjerrum-Bohr \emph{et al.} showed that long-range quantum corrections to gravity arise from vacuum polarization effects of massless particles \cite{Bjerrum-Bohr:2002gqz}. For spherically symmetric black holes, these one-loop quantum corrections manifest as power-law $r^{-3}$ modifications to the macroscopic metric (e.g.,\cite{Xiao:2021zly,Calmet:2021lny}), effectively acting as perturbative quantum hair. Although suppressed by the Planck scale and therefore extremely weak in astrophysical contexts, such corrections may accumulate observable effects during the long-term evolution of Extreme Mass Ratio Inspirals (EMRIs) (e.g.,\cite{Ryan:1995wh,Glampedakis:2005cf,Gair:2012nm}).

EMRIs—systems in which a stellar-mass compact object spirals into a supermassive black hole—serve as ideal probes of such subtle geometric deviations \cite{LISA:2017pwj}. Unlike comparable-mass mergers, EMRIs undergo tens of thousands of orbital cycles within the detector band prior to plunge \cite{Barack:2003fp}. This prolonged evolution enables the small body to perform precision spacetime mapping (geodesy) of the central black hole (e.g.,\cite{Ryan:1995wh,Glampedakis:2005cf}). Even tiny deviations from the classical metric, whether arising from modified gravity or quantum effects, can accumulate over time and produce measurable phase dephasing in gravitational waveforms \cite{Gair:2012nm}. Therefore, monitoring EMRI signals with future space-based gravitational-wave detectors such as LISA, Taiji, and TianQin offers a promising avenue to test quantum hair and strong-field gravity (e.g.,\cite{Babak:2017tow,TianQin:2015yph,Hu:2017mde}).

In this work, we aim to establish a theoretical connection between microscopic quantum corrections and macroscopic gravitational-wave observables. Rather than introducing phenomenological parameters by hand, we adopt the Barvinsky–Vilkovisky covariant effective action formalism (e.g.,\cite{Barvinsky:1985an,Barvinsky:1990up}) and derive the vacuum quantum-corrected black hole metric from a discrete dust shell model in the point-particle limit (e.g.,\cite{Calmet:2019eof,Giacchini:2020dhv}). Building on recent advances in periodic orbit analysis (e.g.,\cite{Levin:2008mq,Healy:2009zm,Barausse:2008xv,Lukes-Gerakopoulos:2012qpc,Drasco:2003ky}), we employ Levin's orbital classification scheme to systematically investigate the dynamics of periodic orbits, particularly the zoom–whirl trajectories that are highly sensitive to strong-field geometry \cite{Healy:2009zm}. Furthermore, using the Numerical Kludge approach (e.g.,\cite{Barack:2003fp,Babak:2006uv}), we model gravitational-wave emission and analyze the cumulative phase effects induced by the quantum hair parameter $\gamma$, providing a model-independent framework for testing quantum gravity signatures through gravitational-wave observations (e.g.,\cite{Gair:2012nm,Yunes:2013dva}).

The structure of the paper is summarized as follows. Sec.~\ref{sec:level2} introduces the vacuum quantum-corrected black hole metric derived from the microscopic dust-shell model and discusses the physical interpretation of the parameters together with the horizon structure. In Sec.~\ref{sec:level3} we investigate the motion of test particles in the quantum-corrected spacetime and examine the properties of bound orbits, with particular emphasis on the marginally bound orbit (MBO) and the innermost stable circular orbit (ISCO). Sec.~\ref{sec:level4} explores how the presence of quantum hair influences the topology of orbital motion within the periodic-orbit classification framework. In Sec.~\ref{sec:level5} we evaluate the corresponding gravitational-wave signals using the Numerical Kludge approach and analyze the resulting phase evolution. Finally, Sec.~\ref{sec:level6} presents the main conclusions and discusses possible future directions.

\section{\label{sec:level2}Vacuum Quantum-Corrected Black Hole Metric Based on Effective Field Theory}

First, consider the quantum-corrected metric describing the exterior spacetime of an $N$-layer equal-mass dust shell \cite{Xiao:2021zly}: 

\begin{widetext}
	\begin{equation}
		\begin{aligned}
			ds^2 = -\left\{1 - \frac{2GM}{r} - \frac{128\pi(\alpha+\beta)G^2M}{r^3}
			\left[1+\frac{R^2(3N^2+3N-1)}{5N^2r^2}\right]\right\} dt^2\\ 
			+ \left[1+\frac{2GM}{r}-\frac{4G^2M^2}{r^2}
			-\frac{384\pi\alpha G^2M}{r^3}
			\left(1+\frac{R^2(3N^2+3N-1)}{5N^2r^2}\right)\right] dr^2 
			+ r^2 d\Omega^2.
		\end{aligned}
		\label{1}
	\end{equation}
\end{widetext}

To obtain a black hole solution without internal structure, we take the point-mass limit, i.e., let the shell radius $R \to 0$. In this limit, the higher-order terms involving $R^2$ (corresponding to the finite-size effect of the matter distribution) tend to vanish. In this expression, the term \( r^2 d\Omega^2 \) describes the angular component associated with spherical symmetry, and the metric can therefore be written in the simplified form shown in Eq.~(\ref{2}):

\begin{widetext}
	\begin{equation}
		\begin{aligned}
			ds^2 &= -\left[1-\frac{2GM}{r}-\frac{C_t}{r^3}\right]dt^2
		    &\quad + \left[1+\frac{2GM}{r}-\frac{4G^2M^2}{r^2}-\frac{C_r}{r^3}\right]dr^2
		    &\quad + r^2d\Omega^2,
		\end{aligned}
		\label{2}
	\end{equation}
\end{widetext}
where $C_t = 128\pi(\alpha+\beta)G^2M$ and $C_r = 384\pi\alpha G^2M$. This form reveals the general scaling behavior of the quantum correction terms at large distances as $1/r^3$ .

While Eq.~(\ref{2}) encodes the algebraic structure associated with a point-like source, its effective coefficients $\alpha$ and $\beta$ must be redefined in order to represent a vacuum black hole consistently. For the dust shell scenario, the analysis exploits the non-local Gauss–Bonnet theorem (e.g.,~\cite{Xiao:2021zly,Calmet:2019eof}), allowing the Riemann-tensor contribution to be incorporated into the Ricci-sector coefficient. In the vacuum case, however, the leading quantum effects stem instead from the non-local operator $\mathcal{R}\ln(\Box/\mu^2)\mathcal{R}$ (e.g.,~\cite{Barvinsky:1985an,Calmet:2019eof}).

Within the Barvinsky–Vilkovisky framework, the renormalization procedure in a vacuum background requires that the coefficients originally tied to the matter sector be reformulated in terms of a single universal parameter $\gamma$, which encapsulates the effect of vacuum fluctuations. 
The parameter $\gamma$ effectively encodes the Wilson coefficient associated with the non-local quantum correction. By enforcing the trace anomaly of the energy–momentum tensor together with the covariant conservation condition (e.g.,~\cite{Calmet:2019eof,Giacchini:2020dhv}), the relation connecting the original coefficients to $\gamma$ can be fixed as

\begin{itemize}
	\item For the time component correction term: $\frac{512\pi\gamma G^2M}{3}$ 
	\item For the spatial component correction term: $-256\pi\gamma G^2M$ 
\end{itemize}
Substituting the above coefficient correspondences into Eq.~(\ref{1}), the quantum-corrected spacetime line element of the vacuum black hole is finally derived \cite{Xiao:2021zly}: 

\begin{equation}
	ds^2 = -A(r)dt^2 + \frac{1}{B(r)}dr^2 + r^2d\Omega^2.
	\label{3}
\end{equation}
Here, $\gamma$ is the parameter describing the strength of the black hole hair. Where $A(r) = 1-\frac{2GM}{r}-\frac{512\pi\gamma G^2M}{3r^3}$ and $B(r) = 1-\frac{2GM}{r}-\frac{256\pi\gamma G^2M}{r^3}$. (Unless otherwise specified, natural units are used in this paper, i.e., $G=1$) .

\section{\label{sec:level3}Orbital Dynamics in a Quantum-Corrected Black Hole Background}

Before turning to the detailed analysis of orbital evolution, one must first identify the physically viable range of the quantum hair parameter. As the radial coordinate decreases, the quantum correction becomes increasingly significant in the strong-field region, and its magnitude consequently plays a decisive role in shaping the spacetime structure, including the possible emergence of horizons.
An extremal black hole satisfies the following degenerate horizon condition:

\begin{equation}
	\begin{cases}
		B(r) = 0 \\
		\left. \dfrac{dB(r)}{dr} \right|_{r=r_h} = 0
	\end{cases}.
	\label{4}
\end{equation}
The extremal black hole configuration corresponds to the critical value of the quantum hair parameter $\gamma = \gamma_c$.Solving Eq.~(\ref{4}) yields $r_h = \frac{4}{3}M$. Substituting this radius back into the horizon equation $B(r_h) = 0$, we can solve for the critical quantum hair parameter corresponding to the extremal black hole:

\begin{equation}
	\gamma_c = -\frac{1}{216\pi} \approx -0.00147.
	\nonumber
\end{equation}

\begin{figure}
	\centering
	\includegraphics[width=0.45\textwidth]{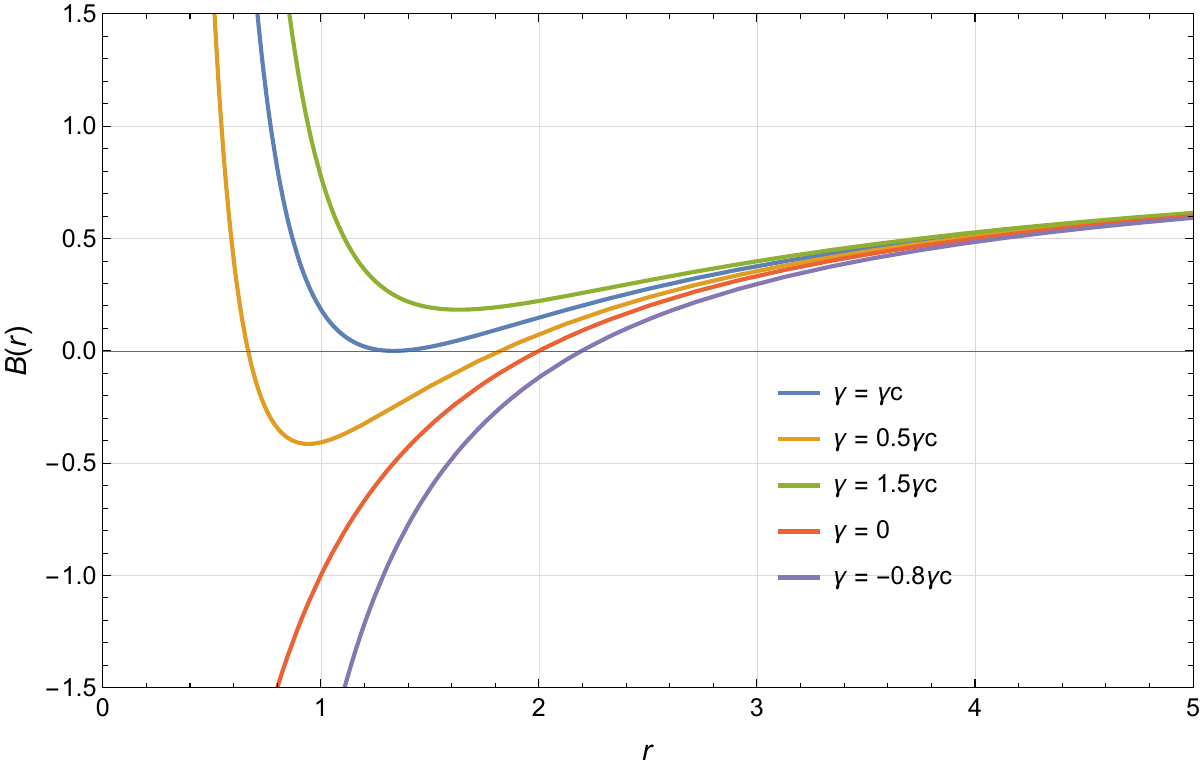}
	\caption{Radial behavior of the metric function $B(r)$ for different values of the quantum hair parameter $\gamma$. The event horizon corresponds to the root of $B(r)=0$.}
	\label{a}
\end{figure}

Fig.~\ref{a} illustrates the behavior of the black-hole metric function together with the associated horizon structure for a range of values of the quantum hair parameter. When $\gamma = 0$ (red curve), the solution reduces to the standard Schwarzschild black hole. When $\gamma>0$ (for instance the purple curve $\gamma=-0.8\gamma_c$, noting that $\gamma_c<0$), the quantum correction effectively enhances the gravitational attraction, causing the event horizon radius to exceed the Schwarzschild value. By contrast, when $\gamma < 0$ the quantum vacuum polarization introduces a repulsive contribution that decreases the horizon radius. As the parameter approaches the critical value $\gamma = \gamma_c$ (blue curve), the inner and outer horizons merge, corresponding to an extremal black-hole configuration. If $\gamma < \gamma_c$ (for instance the green curve $\gamma = 1.5\gamma_c$), the metric function remains positive everywhere, implying that no physical horizon exists and the spacetime develops a naked singularity. These features clearly demonstrate how the quantum parameter modifies the spacetime geometry of the black hole. To maintain the internal consistency of the physical setup considered here, the following analysis will be restricted to the phenomenological parameter region where a physical horizon is present, namely $\gamma \geq \gamma_c$.

With the quantum-corrected spacetime geometry specified, we proceed to formulate the equations of motion for a test particle and analyze the properties of its bound trajectories in the strong-field regime.

We study the motion of a test particle with rest mass $\mu$ moving in the spacetime of a static and spherically symmetric black hole endowed with quantum hair. Because the geometry possesses the timelike Killing vector $\xi_t = \partial_t$ and the axial Killing vector $\xi_\phi = \partial_\phi$, two conserved quantities arise from the corresponding symmetries according to Noether’s theorem: the conserved energy per unit mass E and the specific angular momentum $L$:

\begin{equation}
	E = A(r)\frac{dt}{d\tau},
	\label{5}
\end{equation}

\begin{equation}
	L = r^2\frac{d\phi}{d\tau}.
	\label{6}
\end{equation}
Using the four-velocity normalization condition $g_{\mu\nu}\dot{x}^\mu\dot{x}^\nu = -1$ (the first integral form of the geodesic equation) satisfied by the particle's motion, and expanding it in the equatorial plane ($\theta = \frac{\pi}{2}$):

\begin{equation}
	-A(r)\left(\frac{dt}{d\tau}\right)^2 + \frac{1}{B(r)}\left(\frac{dr}{d\tau}\right)^2 + r^2\left(\frac{d\phi}{d\tau}\right)^2 = -1,
	\label{7}
\end{equation}
we can derive the radial equation of motion for the test particle:

\begin{equation}
	E^2 = \frac{A(r)}{B(r)}\left(\frac{dr}{d\tau}\right)^2 + V_{eff},
	\label{8}
\end{equation}
where the effective potential $V_{eff}(r, \gamma, L)$, corrected by the quantum hair, is defined as:
\begin{equation}
	V_{eff} = A(r)\left(1 + \frac{L^2}{r^2}\right).
	\label{9}
\end{equation}
It can be seen that as $r \to \infty$, we have $\displaystyle \lim_{r \to \infty} V_{eff} = 1$. In this case, if $E > 1$, the particle can escape to infinity ($\left. \dot{r}^2 \right|_{r \to \infty} > 0$), which is not discussed here. In this paper, we focus on the condition $E \leq 1$, where the existence of the effective potential $V_{eff}$ may confine the particle's motion within a certain region.

\section{\label{sec:level4}Critical Orbital Features and Parameter Space Analysis}

With the quantum-corrected geometry and its physically admissible parameter domain established, we proceed to examine the dynamical boundaries arising in the strong-gravity regime. The test-particle system features two characteristic radial thresholds. The first corresponds to the critical potential barrier that separates bound motion from unbound trajectories, namely the marginally bound circular orbit (denoted by $r_{MBO}$), defined by the condition $E = 1$. This orbit marks the outermost boundary of the strong-field region beyond which a black hole can no longer capture infalling matter.

\begin{figure*}[tbph]
	\centering
	\includegraphics[width=1\textwidth]{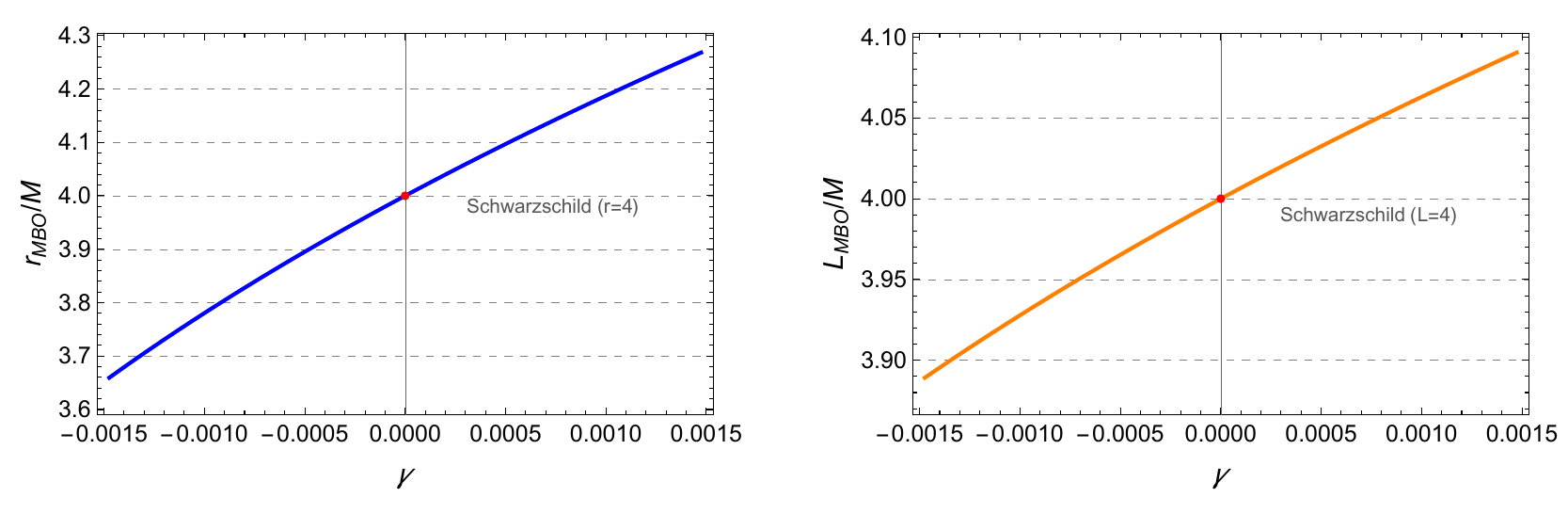}
	\caption{Behavior of the marginally bound orbit radius and the corresponding angular momentum as functions of the quantum hair parameter}
	\label{b}
\end{figure*}

Formally, the parameters of the marginally bound orbit are obtained from the following set of critical conditions:
\begin{equation}
	\begin{cases}
		V_{eff}=E^{2}=1 \\[4pt]
		\dfrac{dV_{eff}}{dr}=0
	\end{cases}
	\tag{10}
	\label{10}
\end{equation}
Owing to the $r^{-3}$ higher-order nonlinear term arising from the quantum corrections in the effective potential $V_{eff}$, the above system does not admit a closed-form analytical expression for $r_{MBO}$. We therefore adopt a numerical method. Through numerical calculations, we investigate how $r_{MBO}$ and $L_{MBO}$ change with $\gamma$, as illustrated in Fig.~\ref{b}. 

The results show that as $|\gamma|$ increases, both $r_{MBO}$ and $L_{MBO}$ decrease monotonically. When $\gamma$ → 0, corresponding to the absence of quantum corrections, these quantities smoothly approach their classical Schwarzschild values. This trend suggests that the quantum-induced repulsive contribution to the effective potential allows marginally bound configurations to occur at smaller orbital radii, i.e., closer to the event horizon in the strong-field regime.

The second characteristic boundary is given by the innermost stable circular orbit, $r_{ISCO}$, which defines the endpoint of stable inspiral motion. Once this critical radius is crossed, the orbit loses stability, and the particle transitions from adiabatic evolution to a rapid radial plunge.

For the innermost stable circular orbit, its parameters are determined by the following system of critical equations:

\begin{equation}
	\begin{cases}
		V_{eff} = E^2 \\[4pt]
		\dfrac{dV_{eff}}{dr} = 0 \\[6pt]
		\dfrac{d^2V_{eff}}{dr^2} = 0
	\end{cases}.
	\tag{11}
	\label{11}
\end{equation}

\begin{figure*}[tbph]
	\centering
	\includegraphics[width=1\textwidth]{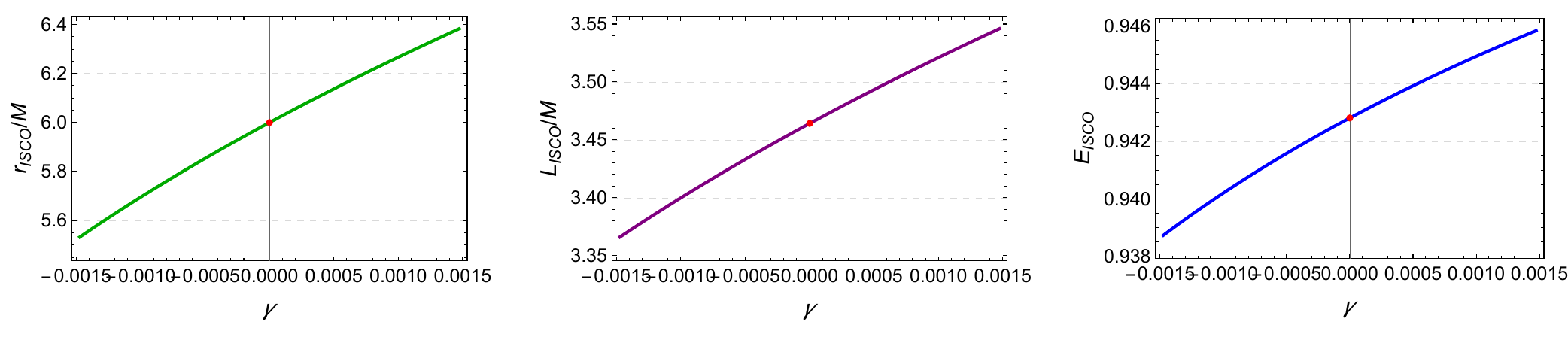}
	\caption{Variation of the ISCO radius and its angular momentum with the quantum hair parameter.}
	\label{c}
\end{figure*}

Following the same numerical solution method as for the marginally bound orbit, we obtain the variations of $r_{ISCO}$, $L_{ISCO}$, and $E_{ISCO}$ with  $\gamma$. The dynamical numerical solutions in Fig.~\ref{c} intuitively reflect the strength of the quantum repulsive effect: compared with the pure Schwarzschild background, the introduction of  $\gamma$ pushes the innermost stable circular orbit to a deeper spacetime layer. This implies that before the irreversible radial plunge occurs, the probe celestial body can maintain adiabatic evolution in the extremely strong-field region much closer to the horizon, and its critical radius, orbital angular momentum, and intrinsic binding energy are all significantly suppressed.

\begin{figure*}[tbph]
	\centering
	\includegraphics[width=1\textwidth]{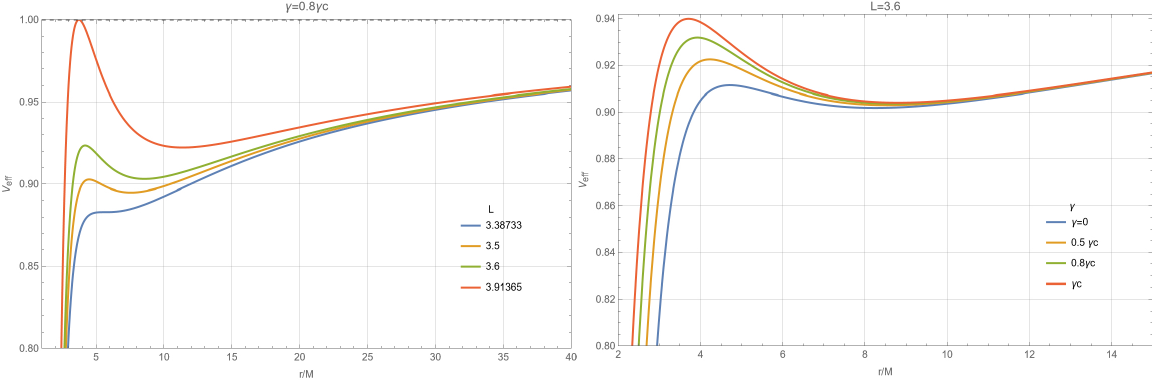}
	\caption{Effective potential distributions under different angular momenta and quantum hair parameters.}
	\label{d}
\end{figure*}

The radial profile of $V_{eff}$ provides a useful way to examine how quantum hair influences the orbital dynamics of test particles. Fig.~\ref{d} displays the effective potential for several representative values of the angular momentum.

The left panel illustrates the variation of $V_{eff}$ with the radial coordinate for several values of $L$, with $\gamma$ fixed at $0.8\gamma_c$. The blue curve corresponds to $L_{ISCO}=3.34344$, for which only one extremum appears. The other blue curve corresponds to $L_{MBO}=3.86321$, exhibiting two extrema. The right panel presents the radial dependence of $V_{eff}$ for different values of $\gamma$ with $L=3.6$ fixed (where $L_{ISCO}<L<L_{MBO}$). It is found that different values of $\gamma$ do not affect the existence of extrema in the curves.

From the trends of these curves, we conclude that the quantum parameter $\gamma$ modifies the curvature of the effective potential in the strong-field region. This modification significantly enhances the binding energy and radiation efficiency of compact objects near the ISCO and induces a drift in the orbital frequency, causing the orbital precession to deviate from the predictions of classical general relativity (e.g.,\cite{Ryan:1995wh,Glampedakis:2005cf}). Consequently, detectable quantum-correction fingerprints are imprinted in the gravitational-wave phase.

\begin{figure}[tbph]
	\centering
	\includegraphics[width=0.45\textwidth]{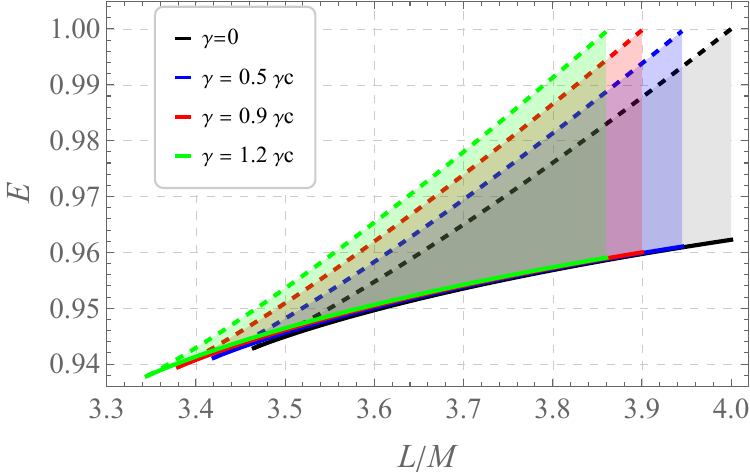}
	\caption{Allowed orbital regions for particles under different quantum hair parameters.}
	\label{e}
\end{figure}

Fig.~\ref{e} illustrates the region of the $E$-$L$ parameter space that allows bound particle motion for different values of the quantum hair parameter, where the black curve represents the Schwarzschild solution, the accessible region of the particle deviates from the Schwarzschild case as the quantum parameter decreases.

Combining the four-velocity normalization condition (\ref{7}) and the conserved definitions of energy and angular momentum from the previous section, we can derive the radial equation of motion:

\begin{equation}
	\dot{r}^2 = B(r)\left[\frac{E^2}{A(r)} - \left(1 + \frac{L^2}{r^2}\right)\right].
	\tag{12}
	\label{12}
\end{equation}

\begin{figure}[tbph]
	\centering
	\includegraphics[width=0.45\textwidth]{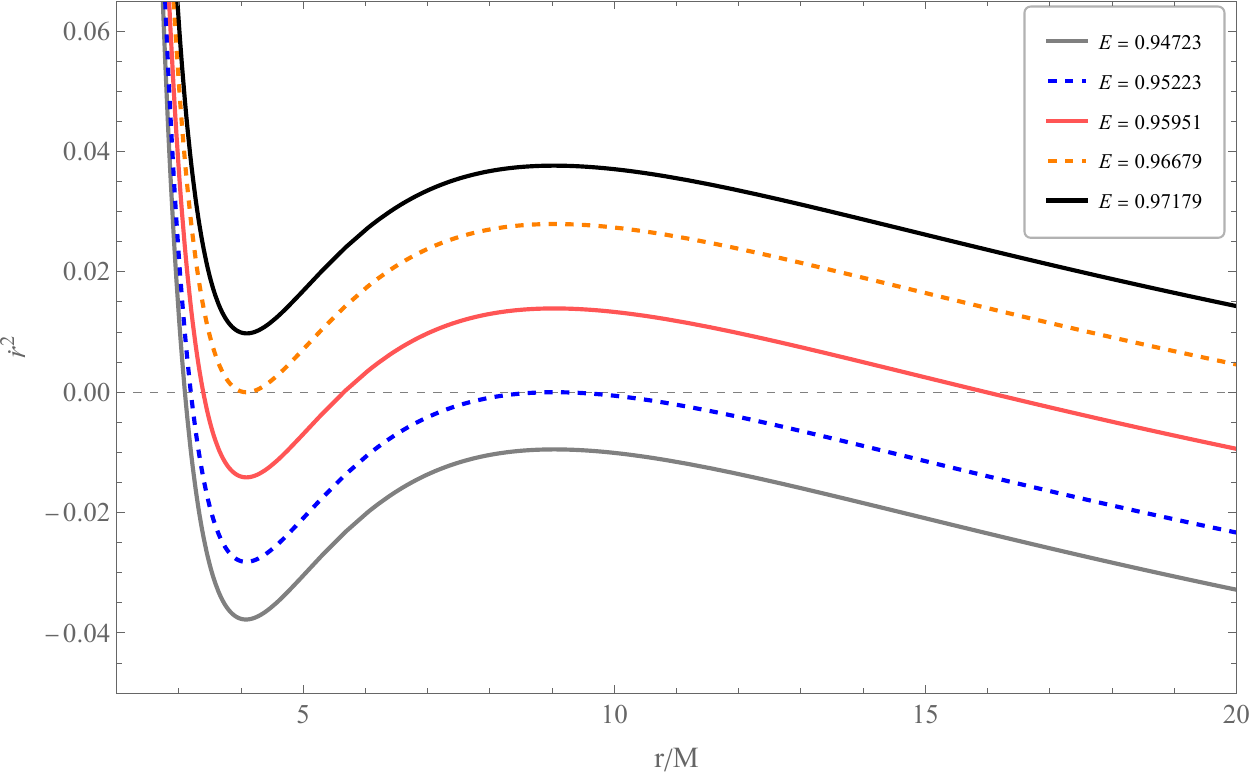}
	\caption{Radial motion function of particles around a quantum-corrected black hole.}
	\label{f}
\end{figure}

Fig.~\ref{f} illustrates the radial behavior of the function $\dot{r}^2$ associated with the particle equation of motion in the quantum black hole spacetime, plotted as a function of the radial coordinate \(r\). In this example the specific angular momentum is fixed at $L = \frac{L_{ISCO} + L_{MBO}}{2} = 3.65048917$. For the quantum hair parameter $0.8\gamma_c$, increasing the particle energy $E$ shifts the curve upward as a whole. Correspondingly, the number of solutions to the equation $\dot{r}^2 = 0$ evolves in the sequence: one root, two roots, three roots, two roots, and finally one root. These features correspond to different bound states of the particle motion.

For $\gamma = 0.8\gamma_c$, increasing $E$ raises the $\dot{r}^2$ curve and changes the number of solutions of $\dot{r}^2=0$ successively from one to two, then to three, and finally back to two and one. This sequence reflects the corresponding changes in the orbital motion of the particle. For $E<0.94574$, no bound motion exists and the particle plunges directly into the black hole. At the critical energy $E=0.95074$, although two solutions appear, the interval between them corresponds to $\dot{r}^2<0$, so bound orbits are still not allowed.

Bound trajectories emerge once the particle energy enters the interval $0.95074 < E < 0.96576$. In this regime the equation $\dot r^{2}=0$ possesses three real roots, which define a physically allowed radial region between the two larger roots where $\dot r^{2}>0$. This feature indicates the presence of bound particle motion and is consistent with the allowed $E$-$L$ parameter region shown in Fig.~\ref{e}. As the energy increases to $E$=$0.96576$, the inner pair of roots coalesces, while bound motion remains possible within the radial range determined by the remaining two roots. When the energy further reaches $E$=$0.97076$, the condition $\dot{r}^2 > 0$ is satisfied only within a limited radial interval, implying that the particle eventually plunges into the black hole from a finite radius rather than from spatial infinity.

\section{\label{sec:level5}Classification of Periodic Orbits in a Quantum-Corrected Background}

In EMRIs, the motion of a compact secondary object in the strong gravitational field of a massive black hole provides a sensitive probe of the underlying spacetime geometry. In this section we employ the periodic-orbit classification scheme introduced by Levin and Perez-Giz (e.g.,\cite{Levin:2008mq,Healy:2009zm}). This framework characterizes black hole orbits through a rational number $q$, allowing a quantitative description of how the quantum-correction parameter $\gamma$ influences the topological structure of the trajectories. The relation between the rational number $q$ and the orbital triplet is defined as:

The rational number $q$ associated with the periodic orbit can be written in terms of the triplet as:
\begin{equation}
	q = \frac{\Delta \Phi}{2\pi} - 1 = w + \frac{v}{z},
	\tag{13}
	\label{13}
\end{equation}
here $\Delta \Phi$ denotes the total azimuthal angle accumulated by the particle during a full radial oscillation, namely from apastron to periastron and back again. The integer $z$ specifies the number of zoom segments, $w$ represents the whirl number, and $v$ characterizes the vertex index of the periodic orbit.

Restricting the motion to the equatorial plane ($\theta = \frac{\pi}{2}$), the orbital equation can be obtained by combining the angular momentum relation $\frac{d\phi}{d\tau} = \frac{L}{r^2}$ with the radial equation of motion (\ref{11}), thereby removing the affine parameter $\tau$. This leads to an equation for $\frac{d\phi}{dr}$. For the quantum-corrected black hole metric introduced in this work, the expression for the azimuthal angle accumulation \(\Delta\Phi\) must include the modified metric functions $A(r)$ and $B(r)$. In particular, the azimuthal advance between the radial turning points $r_a$ and $r_b$ is written as:
\begin{equation}
	\Delta \Phi = \oint d\phi = 2 \int_{r_a}^{r_b} \left| \frac{d\phi}{dr} \right| dr,
	\nonumber
\end{equation}

\begin{equation}
	\Delta \Phi = 2 \int_{r_a}^{r_b} \frac{L}{r^2} \sqrt{\frac{A(r)}{B(r)(E^2 - V_{eff})}} dr.
	\tag{14}
	\label{14}
\end{equation}
A numerical evaluation of Eq.~(\ref{14}) reveals how the rational number $q$ depends on $E$ and $L$. If $q$ takes a rational value, the trajectory closes after a finite number of radial oscillations, corresponding to a periodic orbit. In contrast, when $q$ is irrational, the trajectory does not close and instead appears as a precessing orbit.

\begin{figure}
	\centering
	\includegraphics[width=0.45\textwidth]{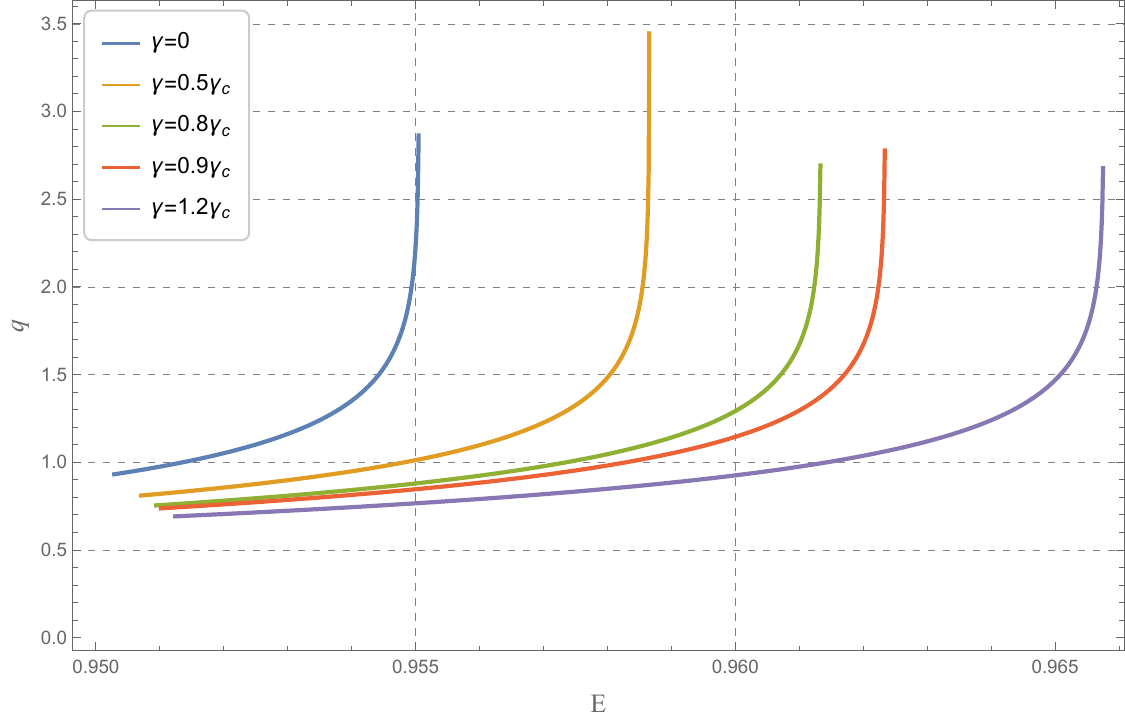}
	\caption{The trend of the rational number q as a function of energy E (with L fixed)}
	\label{g}
\end{figure}

\begin{figure}
	\centering
	\includegraphics[width=0.45\textwidth]{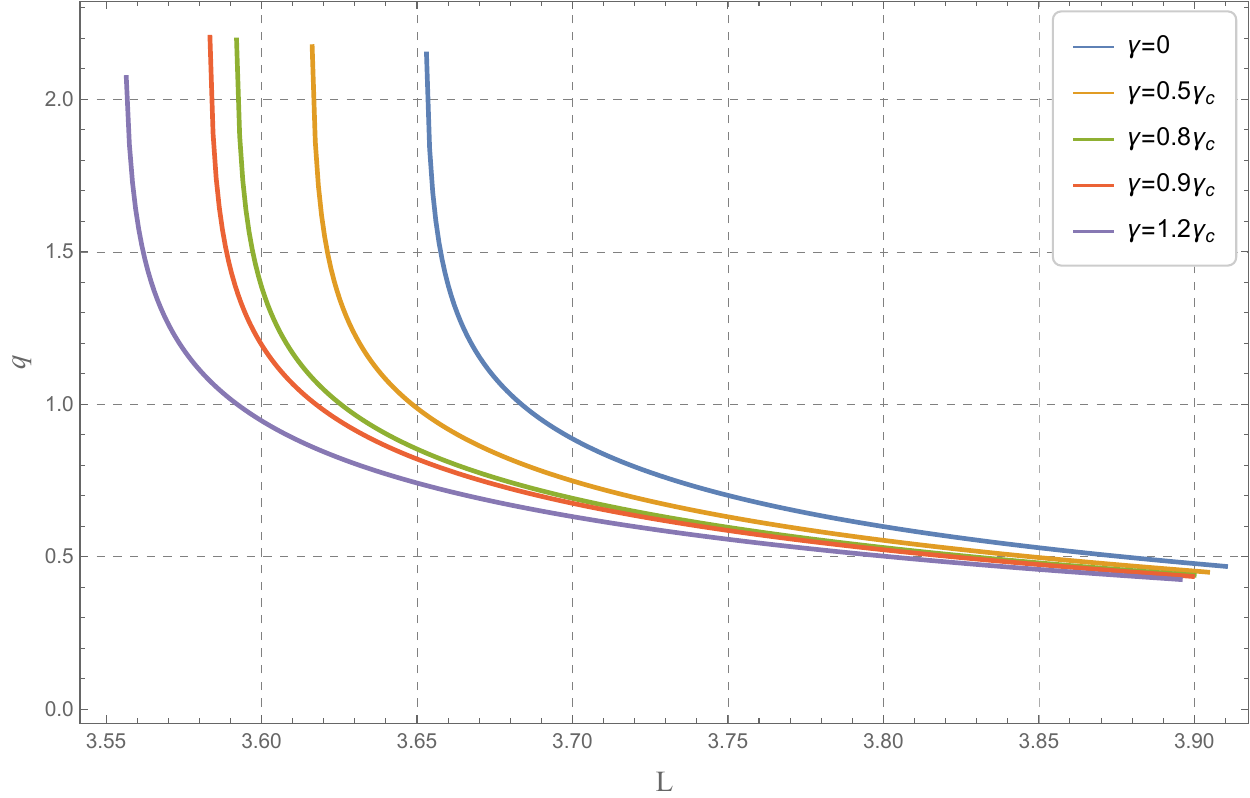}
	\caption{The trend of rational number q with L (fixed E)}
	\label{h}
\end{figure}

Fig.~\ref{g} illustrates the variation of $q$ with $E$ for a fixed value of $L$ (here $L = \frac{L_{ISCO} + L_{MBO}}{2} = 3.65048917$). Fig.~\ref{h}, on the other hand, shows the dependence of $q$ on $L$ while the energy is held fixed at $E = 0.96$.With the aid of this topological tracking, we have captured the nonlinear modulation of the orbital precession frequency by quantum hair. The evolution curves clearly show that when the orbital parameters of the system approach the separatrix, the value of $q$, which represents the ratio of orbital zoom to whirl cycles, undergoes a dramatic distortion.
Combined with Fig.~\ref{d}, we see that the quantum correction term $-1/r^3$ deepens the bottom curvature of the effective potential well. This microscopic mechanism significantly accelerates the aforementioned evolution process, causing the divergence points in orbital dynamics to be triggered earlier at lower thresholds of energy or angular momentum. This explains why, under the same orbital topological mode, the quantum hair black hole forces test particles to maintain stable evolution with lower initial energy and angular momentum, thereby profoundly altering the periodic motion spectrum around the black hole.

Periodic orbits are characterized by the triplet $(z, w, v)$. The corresponding numerical results are summarized in Table~\ref{table1} and Table~\ref{table2}. 
As indicated in Table~\ref{table1}, when $E$ is fixed at $0.96$, the angular momentum $L$ associated with a given periodic orbit decreases as the absolute value of the quantum parameter $\gamma$ increases. In contrast, if $L = \frac{L_{ISCO} + L_{MBO}}{2}$ is adopted, the corresponding energy $E$ for the same periodic orbit also increases with increasing $|\gamma|$. Table~\ref{table2} additionally shows that periodic orbits with larger whirl numbers tend to require slightly smaller values of $L$ or $E$, which agrees with the typical dynamical behavior of zoom--whirl orbits (e.g.,~\cite{Healy:2009zm,Glampedakis:2002ya}).

Relative to the classical Schwarzschild case, the inclusion of the quantum parameter $\gamma$ leads to smaller values of $L$ for fixed $E$, while for fixed $L$ the required orbital energy $E$ increases slightly.

\begin{table*}[]
	\centering
	\begin{tabular}{ccccccccc}
		\hline\hline
		\rule{0pt}{12pt}
		$\gamma$ & $L_{(1,1,0)}$ & $L_{(1,2,0)}$ & $L_{(1,2,1)}$ & $L_{(2,1,1)}$ & $L_{(2,2,1)}$ & $L_{(3,1,1)}$ & $L_{(3,1,2)}$ & $L_{(4,2,1)}$ \\
		\hline
		\rule{0pt}{11pt}
		$0$ & 3.68357627 & 3.65340503 & 3.65258078 & 3.65759304 & 3.65270050 & 3.66172534 & 3.65533281 & 3.65290600 \\
		\rule{0pt}{11pt}
		$0.5\gamma_c$ & 3.64841743 & 3.61670446 & 3.61581158 & 3.62116655 & 3.61594288 & 3.62553348 & 3.61876595 & 3.61616618 \\
		\rule{0pt}{11pt}
		$0.8\gamma_c$ & 3.62545721 & 3.59245551 & 3.59149749 & 3.59716100 & 3.59164016 & 3.60172860 & 3.59463737 & 3.59188053 \\
		\rule{0pt}{11pt}
		$0.9\gamma_c$ & 3.61743951 & 3.58392547 & 3.58293986 & 3.58873090 & 3.58308746 & 3.59337898 & 3.58615727 & 3.58333508 \\
		\rule{0pt}{11pt}
		$1.2\gamma_c$ & 3.59212941 & 3.55674748 & 3.55565303 & 3.56193113 & 3.55582053 & 3.56687663 & 3.55916959 & 3.55609682 \\
		\hline\hline
	\end{tabular}
	\caption{The angular momentum $L$ of periodic orbits under different $(z, w, v)$ and quantum parameter $\gamma$ (with fixed energy $E = 0.96$).}
	\label{table1}
\end{table*}

\begin{table*}[]
	\centering
	\begin{tabular}{ccccccccc}
		\hline\hline
		\rule{0pt}{12pt}
		$\gamma$ & $E_{(1,1,0)}$ & $E_{(1,2,0)}$ & $E_{(1,2,1)}$ & $E_{(2,1,1)}$ & $E_{(2,2,1)}$ & $E_{(3,1,1)}$ & $E_{(3,1,2)}$ & $E_{(4,2,1)}$ \\
		\hline
		\rule{0pt}{11pt}
		$0$ & 0.96542635 & 0.96838279 & 0.96844186 & 0.96802653 & 0.96843431 & 0.96764411 & 0.96822500 & 0.96842007 \\
		\rule{0pt}{11pt}
		$0.5\gamma_c$ & 0.96647823 & 0.96998139 & 0.97005606 & 0.96954643 & 0.97004625 & 0.96908779 & 0.96356671 & 0.96384502 \\
		\rule{0pt}{11pt}
		$0.8\gamma_c$ & 0.96751244 & 0.97154411 & 0.97163693 & 0.97102620 & 0.97162434 & 0.97049064 & 0.96032155 & 0.96068474 \\
		\rule{0pt}{11pt}
		$0.9\gamma_c$ & 0.96796294 & 0.97223109 & 0.97233308 & 0.97167351 & 0.97231902 & 0.97110275 & 0.95914576 & 0.95954783 \\
		\rule{0pt}{11pt}
		$0.2|\gamma_c|$ & 0.96515116 & 0.96795544 & 0.96801065 & 0.96761990 & 0.96800364 & 0.96725852 & 0.96988183 & 0.97005470 \\
		\hline\hline
	\end{tabular}
	\caption{The energy $E$ of periodic orbits under different $(z, w, v)$ and quantum parameter $\gamma$ (with fixed angular momentum $L=\frac{L_{\rm ISCO}+L_{\rm MBO}}{2}$).}
	\label{table2}
\end{table*}

\begin{figure*}
	\centering
	\includegraphics[width=1\textwidth]{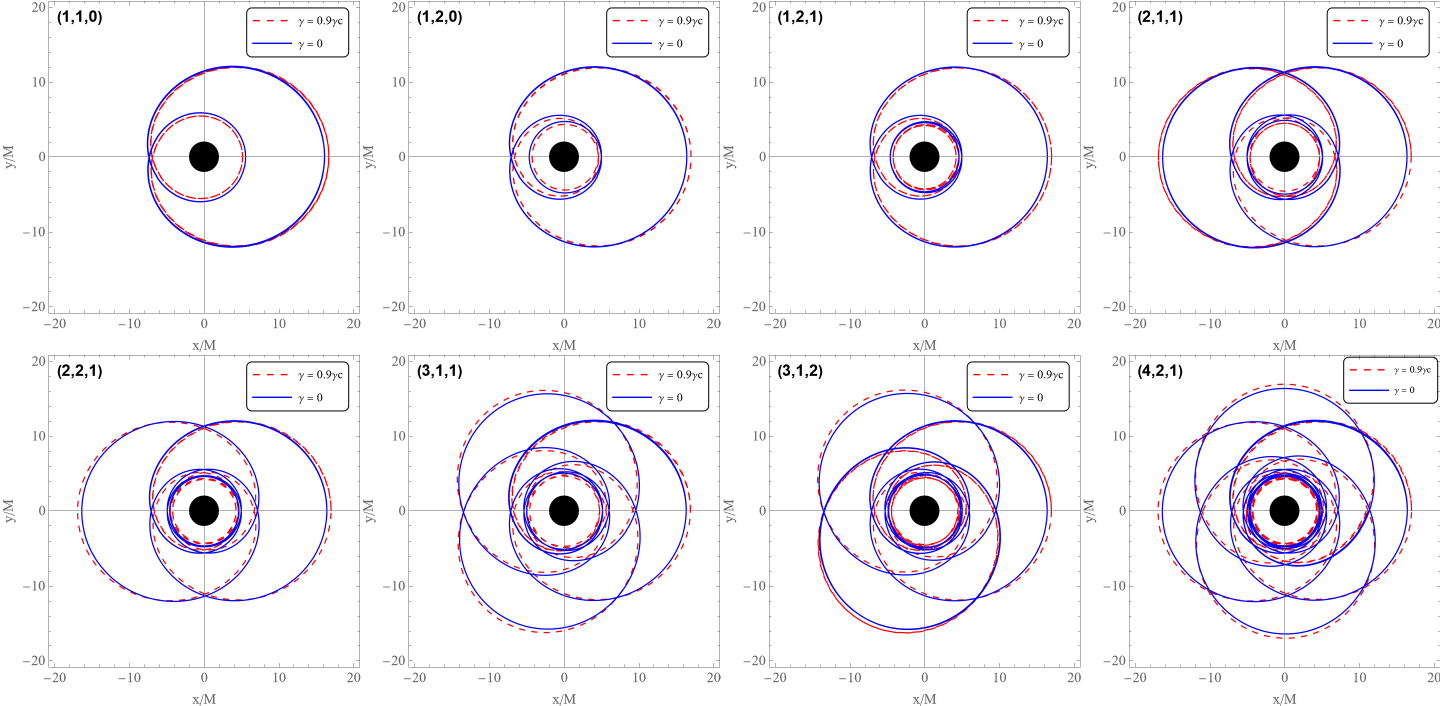}
	\caption{Periodic orbital trajectories of test bodies around a quantum-corrected black hole (fixed $E$)}
	\label{i}
\end{figure*}

After obtaining the data in the tables, we use it to plot the trajectory diagrams for $(z, w, v)$. Fig.~\ref{i} shows a comparison of the trajectories obtained from the table for $L$ at $\gamma = 0.9\gamma_c$ with fixed $E = 0.95$, against the Schwarzschild case. The comparison results intuitively reveal the quantum correction effect: although the orbits under the corrected gravity still maintain a closed, periodic topological structure, the quantum parameter $\gamma$ has modified the effective potential Eq.~(\ref{9}) of spacetime. This causes the red dashed trajectory to spatially separate from the classical Schwarzschild orbit: the inner orbit (periastron) is closer to the black hole, while the outer orbit (apastron) is farther away from it.

\begin{figure}
	\centering
	\includegraphics[width=0.45\textwidth]{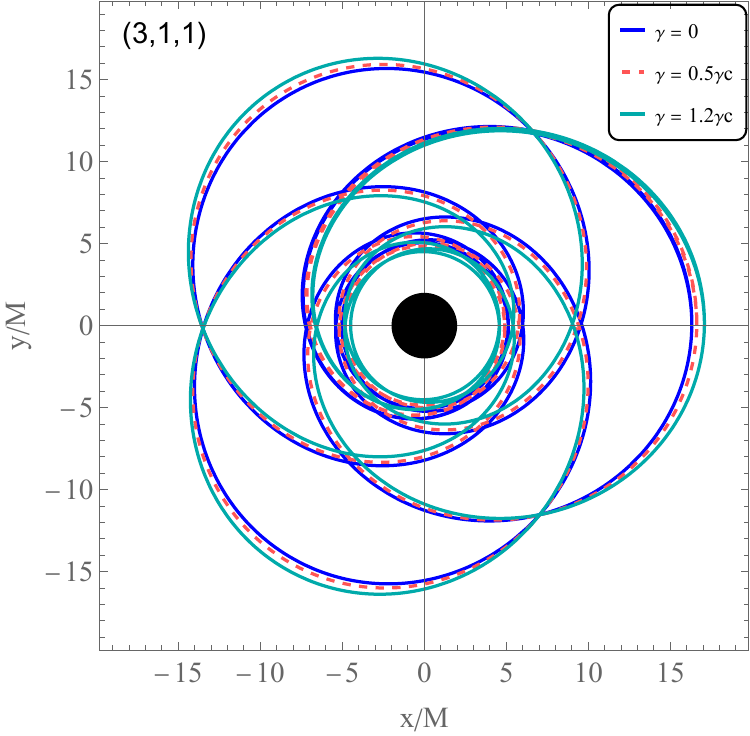}
	\caption{Test-body trajectories under different quantum parameters (fixed energy)}
	\label{j}
\end{figure}

Fig.~\ref{j} shows the trajectories of the test particle for several values of the quantum parameter $\gamma$. As $\gamma$ increases, the periapsis of the orbit shifts toward the black hole while the apoapsis moves outward. This behavior becomes increasingly pronounced relative to the Schwarzschild case.

In summary, quantum hair not only alters the parameter space of bound orbits (the $E$-$L$ space) but also significantly modulates the periodic motion trajectories of test particles. These subtle deviations in orbital dynamical characteristics will inevitably leave a unique "fingerprint" in the gravitational wave waveforms radiated by the particles. To quantitatively investigate this effect, it is essential to numerically compute the specific gravitational wave radiation waveforms and their spectral characteristics.

\section{\label{sec:level6}Gravitational Wave Radiation from Periodic Orbits in a Quantum-Corrected Black Hole Background}

Extreme mass-ratio inspirals generate gravitational waves that contain rich information about the spacetime geometry of the central black hole.
 In the preceding section we analyzed how the quantum hair parameter modifies the geometric properties of periodic particle orbits. Building on these results, we now investigate the gravitational-wave radiation associated with such periodic trajectories in the quantum-corrected background.Although the background spacetime includes quantum correction terms, given that these terms are short-ranged ($\sim r^{-3}$) and the spacetime is asymptotically flat, we adopt the Numerical Kludge approximation \cite{Babak:2006uv}: we assume that the generation and propagation of gravitational waves are primarily described by the quadrupole formula in flat spacetime, while the effects of quantum hair are manifested in the gravitational wave waveforms mainly through modifying the orbital dynamics of test particles (e.g., orbital frequency drift and periastron precession). Next, we focus on investigating the modulation effect of the quantum hair parameter $\gamma$ on the amplitude, phase, and frequency characteristics of gravitational waves, to assess its observability in future space-based gravitational wave detections.

Specifically, under the Numerical Kludge approximation, we assume that test particles move along geodesics in the real curved spacetime background (the quantum-corrected black hole metric described by Eq.~(\ref{3})), while the gravitational radiation field they emit is approximately calculated using the multipole moment formula in flat spacetime. Since quadrupole radiation dominates the energy flux of gravitational waves, this paper primarily considers the contribution of the mass quadrupole moment. To convert the dimensionless orbital dynamics solutions into gravitational wave strains with realistic astrophysical magnitudes, we construct the following computational framework:

We project the spherical coordinate trajectory $(r(t), \phi(t))$ of the test particle on the equatorial plane of the black hole onto a virtual flat Cartesian coordinate system:

\begin{equation}
	x(t) = r(t) \cos \phi(t), \quad y(t) = r(t) \sin \phi(t), \quad z(t) = 0.
	\tag{15}
	\label{15}
\end{equation}
Here, $r(t)$ and $\phi(t)$ are the precise solutions obtained by numerically integrating the aforementioned geodesic Eq.~(\ref{12}). To evaluate the gravitational-wave signal we employ the Einstein quadrupole approximation.In the weak-field and slow-motion limit, the spatial part of the metric perturbation $h_{ij}$ measured at a luminosity distance $D_L$ is determined by the second time derivative of the system's mass quadrupole tensor $I_{ij}$ \cite{Peters:1963ux}:

\begin{equation}
	h_{ij}^{TT}(t) = \frac{2G}{c^4 D_L} \frac{d^2 I_{ij}^{TT}}{dt^2} \approx \frac{2G\mu}{c^4 D_L} \left(2v_i v_j + a_i x_j + a_j x_i\right),
	\tag{16}
	\label{16}
\end{equation}
where $\mu$ denotes the mass of the test particle, while $v_i$ and $a_i$ represent the velocity and acceleration components of the particle in the locally flat (virtual Minkowski) frame, respectively. The prefactor $2\mu/D_L$ appearing in the waveform expression plays an essential role in relating the theoretical prediction to observable quantities, converting the dimensionless orbital variables (in geometric units) into physical gravitational-wave signals.
In our numerical setup, we take the mass of the central supermassive black hole to be $M = 10^{6}M_{\odot}$ and the mass of the test body to be $\mu = 10M_{\odot}$. The luminosity distance is chosen as $D_L = 2\,\mathrm{Gpc}$. To model the signal detected by an observer, the metric perturbation $h_{ij}$ calculated in the source frame is projected onto the observer's radiation frame (TT gauge) \cite{Misner:1973prb}.The inclination angle $\iota$ is defined as the angle between the line of sight and the spin axis of the black hole (equivalently, the normal direction of the orbital plane) \cite{Barack:2003fp}. In addition, the observer's azimuthal angle $\zeta$, corresponding to the longitude of periastron, is introduced.After constructing the orthonormal basis associated with the detector frame, the gravitational-wave strain generated by the equatorial orbit (for which $h_{xx}$, $h_{yy}$, and $h_{xy}$ are the only nonvanishing components when $z=0$) is projected onto the polarization basis. The two independent polarization modes $h_{+}$ and $h_{\times}$ are then written as:

\begin{equation*}
	h_+ = \frac{1}{2}\left(h_{\zeta\zeta} - h_{uu}\right),
\end{equation*}
\begin{equation*}
	h_\times = h_{\iota\zeta},
\end{equation*}
where, according to the projection geometry of the strain along the observer's line of sight, the individual projection components are given by:

\begin{align*}
	h_{\zeta\zeta} &= h_{xx} \cos^2 \zeta - h_{xy} \sin 2\zeta + h_{yy} \sin^2 \zeta, \\
	h_{uu} &= \cos^2 \iota \left(h_{xx} \sin^2 \zeta + h_{xy} \sin 2\zeta + h_{yy} \cos^2 \zeta\right), \\
	h_{\iota\zeta} &= \cos \iota \left(\frac{1}{2}h_{xx} \sin 2\zeta + h_{xy} \cos 2\zeta - \frac{1}{2}h_{yy} \sin 2\zeta\right).
	\nonumber
\end{align*}
Combined with Eq.~(\ref{16}), the specific components of the spatial strain are given by the orbital coordinates and dynamical variables of the test particle (e.g.,\cite{Babak:2006uv,Peters:1963ux}):
\begin{align*}
	h_{xx} &= \frac{2G\mu}{c^4 D_L} \left(2\dot{x}^2 + 2x\ddot{x}\right), \\
	h_{yy} &= \frac{2G\mu}{c^4 D_L} \left(2\dot{y}^2 + 2y\ddot{y}\right), \\
	h_{xy} &= \frac{2G\mu}{c^4 D_L} \left(2\dot{x}\dot{y} + x\ddot{y} + y\ddot{x}\right).
\end{align*}

In the subsequent numerical calculations, to facilitate comparison and illustrate the typical features of the gravitational waveforms, we adopt the canonical observation inclination $\iota = \frac{\pi}{4}$ and azimuthal angle $\zeta = \frac{\pi}{4}$.

\begin{figure}
	\centering
	\includegraphics[width=0.45\textwidth]{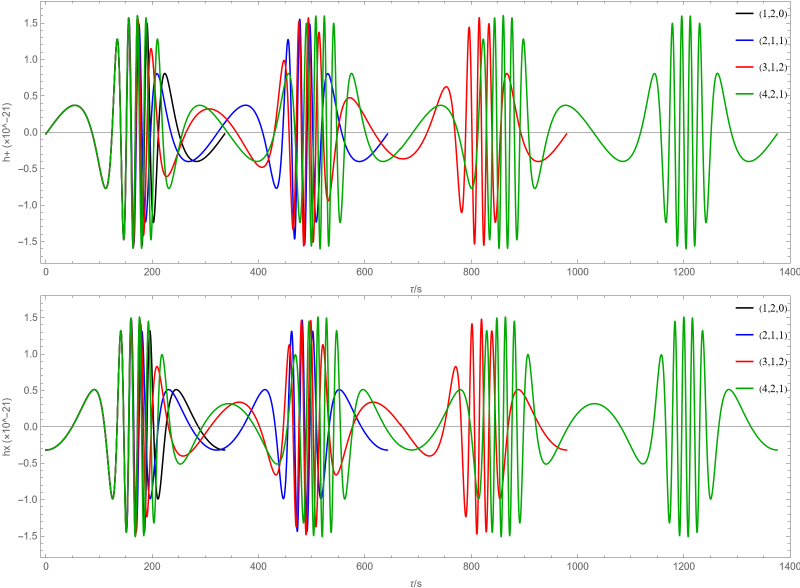}
	\caption{Gravitational-wave signals produced by different periodic orbits at $0.9\gamma_c$. The upper panel displays $h_+$, while the lower panel shows $h_\times$ ($E=0.96$).}
	\label{k}
\end{figure}
From Fig.~\ref{k}, the intermittent waveform structure unique to Zoom-Whirl orbits can be clearly observed. Due to the high eccentricity of the orbit, the particle experiences two distinct phases in its radial motion: 
1) Zoom phase (far-field region): when the particle moves toward apastron, the gravitational potential is weak, and the particle's velocity slows down, resulting in extremely weak gravitational wave radiation, manifested as "quiet periods" or low-frequency, low-amplitude oscillations in the waveform. 
2) Whirl phase (strong-field region): when the particle approaches periastron, it is confined by the strong gravitational field and undergoes multiple rapid revolutions around the black hole. This causes the gravitational wave amplitude to increase dramatically, manifesting as a series of dense high-frequency pulses. 
This unique "chirp-like" signal is the key feature that distinguishes Zoom-Whirl orbits from ordinary elliptical orbits.

\begin{figure*}
	\includegraphics[width=1\textwidth]{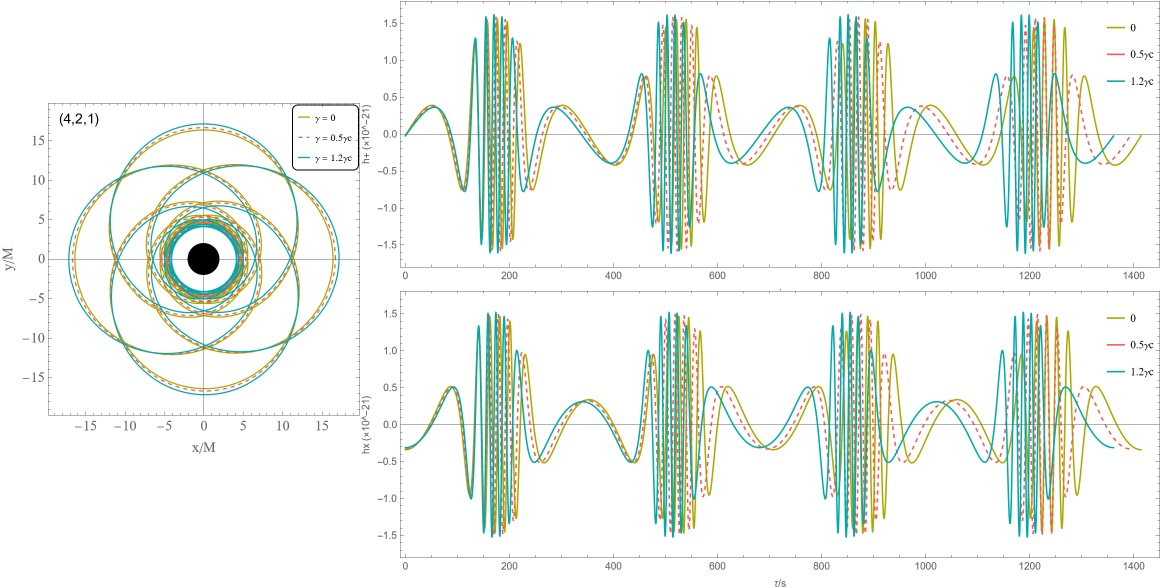}
	\caption{Trajectories and waveform diagrams of the same orbit under different quantum parameters $\gamma$}
	\label{l}
\end{figure*}

Fig.~\ref{l} shows the evolution of gravitational waveforms corresponding to different quantum parameters $\gamma$ under the same initial orbit parameters $(4,2,1)$. In the initial stage of waveform evolution, the waveforms for different $\gamma$ almost coincide. However, as time progresses, significant phase mismatches appear in the waveforms \cite{Gair:2012nm}.

It should be emphasized that since the simulations in this chapter are based on the conservative geodesic equation and do not include the radiation reaction force term, this phase difference does not originate from a change in the rate of orbital decay \cite{Glampedakis:2002ya}. The fundamental physical reason lies in the reconstruction of the spacetime effective potential $V_{eff}(r)$ by the quantum-corrected metric: the $r^{-3}$ correction term directly changes the radial oscillation frequency $\Omega_r$ and the azimuthal precession frequency $\Omega_\phi$ of the test particle in the strong-field region. This slight drift in the fundamental frequency caused by quantum hair accumulates linearly over an evolution of tens of thousands of cycles, eventually leading to obvious misalignment of the wave peaks. This strongly demonstrates that even at the level of conservative dynamics, the quantum hair effect is sufficient to leave a unique observational "fingerprint" in EMRIs signals.

\section{\label{sec:7}Conclusion}

Within the effective field theory framework, we have examined black hole geometries that include vacuum-polarization–induced quantum corrections. Unlike the classical no-hair picture, these power-law contributions modify the gravitational potential outside the horizon and introduce measurable changes in the strong-field region. Our conservative dynamical analysis of extreme mass-ratio inspirals shows that the quantum hair parameter reshapes both the depth and the radial profile of the effective potential. As a result, marginally bound configurations shift inward, and the characteristic radii associated with the MBO and ISCO decrease accordingly. In practical terms, the orbital energy and angular momentum required to maintain stable motion are reduced, leading to a deformation of the accessible phase space.

To explore the structure of bound motion in more detail, we employed a topological index to classify periodic trajectories. The numerical results indicate that when the orbital energy approaches the separatrix, the rational number characterizing the zoom–whirl cycle ratio varies in a strongly nonlinear manner. Quantum corrections enhance this sensitivity. Relative to the Schwarzschild limit, increasing the magnitude of the quantum parameter lowers the energy and angular momentum thresholds needed to sustain a given resonant configuration. This effect is also apparent in configuration space, where both periastron and apastron positions display systematic shifts. In this sense, microscopic geometric modifications are reflected consistently in both phase-space properties and real-space orbital structure.

We next investigate the gravitational radiation associated with representative periodic trajectories in an extreme mass-ratio system. The waveforms were generated using the Numerical Kludge prescription while keeping the quantum hair parameter fixed. The signals reproduce the characteristic burst-like structure associated with zoom–whirl motion. At the same time, the corrected spacetime geometry induces small shifts in the radial and azimuthal fundamental frequencies. Although modest for individual cycles, these frequency differences accumulate over long inspiral durations and lead to a noticeable phase offset in the waveform.

Taken together, these results indicate that the quantum hair parameter leaves subtle but systematic imprints on near-horizon geometry and on the emitted gravitational radiation. Future space-based detectors such as LISA may therefore provide access not only to detailed orbital information in EMRIs, but also to possible quantum deviations from classical black hole spacetimes through high-precision phase measurements. Such observations would offer an avenue to test the no-hair paradigm beyond its purely classical formulation.

\section{acknowledgements}
This work was supported by the Guizhou Provincial Basic Research Program (Natural Science) (Grant No. QianKeHeJiChu[2024]Young166), the National Natural Science Foundation of China (Grant No. 12365008), the Guizhou Provincial Basic Research Program (Natural Science) (Grant No. QianKeHeJiChu-ZK[2024]YiBan027 and QianKeHeJiChuMS[2025]680), the Guizhou Provincial Major Scientific and Technological Program XKBF(2025)·010 (Hosted by Professor Xu Ning), the Guizhou Provincial Major Science and Technological Program XKGF(2025)·009 (Hosted by Professor Xiang Guoyong), and the Guizhou Provincial Major Scientific and Technological Program (Hosted by Teacher Fan Lulu).

%\nocite{*}
%\bibliographystyle{unsrt}
%\bibliography{ref}% Produces the bibliography via BibTeX.

\bibliography{ref}
\bibliographystyle{apsrev4-1}

\end{document}